\documentclass[12pt]{article}
\usepackage{amsmath,amsfonts,amssymb}
\usepackage{amsthm}
\usepackage{graphicx,latexsym}
\usepackage{epsfig}
\usepackage{mathrsfs}
\usepackage{graphicx,epstopdf}
\usepackage{subfigure}
\usepackage{enumitem}
\UseRawInputEncoding
\newcommand{\doublespacing}{\let\CS=\@currsize\renewcommand{\baselinesstrech}
{2.0}\tiny\CS}

\newcommand{\bd}{\begin{document}}
\newcommand{\ed}{\end{document}}
\newcommand{\bc}{\begin{center}}
\newcommand{\ec}{\end{center}}
\newcommand{\bfg}{\begin{figure}}
\newcommand{\efg}{\end{figure}}
\newcommand{\vs}{\vspace}
\newcommand{\beqas}{\begin{eqnarray*}}
\newcommand{\eeqas}{\end{eqnarray*}}
\newcommand{\pad}{\partial}
\newcommand{\f}{\frac}
\DeclareMathOperator{\sech}{sech}

\begin{document}

\title {Formation and interaction of two-dimensional electron-acoustic solitons and breathers in superthermal plasmas }

\author{Jayshree Mondal$^{1,*}$, Prasanta Chatterjee$^1$ and Biswajit Sahu$^{2}$ \\ $^1$Department of
Mathematics, Siksha Bhavana, Visva Bharati University, \\Santiniketan - 731 235, India \\ $^2$Department of
Mathematics, West Bengal State University, \\Barasat, Kolkata-700 126, India}

  \date{}

  \maketitle
  \vspace{0.2cm}
  \centerline{{\bf ABSTRACT}}

  \vspace {0.2 cm}

  \thispagestyle{empty}

  \setlength{\baselineskip}{18.5 pt}
The nonlinear evolution and mutual interaction of two-dimensional electron acoustic (EA) nonlinear structures in superthermal plasma environment are studied. The plasma model consists of inertial cold electrons, superthermal hot electrons described by kappa ($\kappa$) distribution, and stationary ions providing overall charge neutrality. Using the extended Poincar\'e-Lighthill-Kuo (PLK) reductive perturbation technique, a pair of two-sided Kadomtsev-Petviashvili (KP) equations governing right- and left-propagating EA solitary waves (EASWs) is derived. Exact analytical solutions of the KP equations, including single soliton, multisoliton, breather, and lump structures, are obtained via the Hirota bilinear method. The effects of key plasma parameters such as hot electron concentration, temperature ratio, and superthermality index on the characteristics of these nonlinear excitations are examined. Particular attention is devoted to the head-on collision dynamics between solitons, breather-soliton, and breather-breather interactions. The results reveal quasi-elastic collisions accompanied by phase shifts, transient amplitude modulation, and localized energy concentration, with clear distinctions between oscillatory and non-oscillatory mode interactions. The present study provides new insights into multidimensional electron acoustic wave (EAW) dynamics and energy redistribution mechanisms in superthermal space plasmas, with direct relevance to planetary magnetospheric environments such as Saturn's ring region.

\vspace {0.2 cm}

\vspace {0.17 cm}
 $\overline{{^* Corresponding ~ author ~ E-mail: jayshreemondal2512@gmail.com}}$

\newpage

\section{Introduction}
EAWs constitute an important class of high-frequency electrostatic modes in multicomponent plasmas and have attracted considerable attention over the past few decades owing to their relevance in both laboratory and space plasma environments \cite{eawkp1}-\cite{eawkp5}. EAWs have been invoked to describe the electrostatic component of broadband electrostatic noise (BEN) observed in the cusp region of the Earth's bow shock, terrestrial magnetosphere, and the heliospheric termination shock \cite{eawkp8, eawkp9}. Spacecraft observations from missions such as FAST, POLAR, and GEOTAIL have reported localized electrostatic structures consistent with nonlinear EAW activity in the auroral region and the magnetosphere \cite{eawkp10, eawkp12}.
In addition, strong observational evidence for the existence of EAWs has been reported in planetary magnetospheres, especially in the Saturnian system. The coexistence of cool and hot electron populations in Saturn's magnetosphere was first inferred from the Voyager Plasma Science (PLS) observations by Sittler \textit{et al.} \cite{eawkp13}. They identified distinct electron temperature components in the plasma surrounding Saturn. These results were later confirmed and significantly extended by high-resolution measurements from the Cassini Plasma Spectrometer (CAPS), which revealed persistent two-temperature electron distributions over a wide range of radial distances, including the vicinity of Saturn's rings \cite{eawkp14}. Such plasma conditions are highly favorable for the excitation and propagation of EAWs, with cold electrons providing inertia and hot electrons supplying the restoring force.

The nonlinear properties of EAWs are of special significance, since space plasmas are inherently nonlinear and often support the formation of coherent structures such as solitons, double layers, breathers, and localized wave packets. Most theoretical investigations of EAWs have traditionally employed the Boltzmann or Maxwellian distribution to describe the hot electron population. In space plasma environments, a wide range of physical processes, including wave-particle interactions, velocity-space diffusion, particle trapping, and the action of external forces such as electric and magnetic field fluctuations, continuously drive the plasma away from thermodynamic equilibrium. Consequently, the propagation characteristics of plasma waves, as well as the nature of the associated nonlinear structures, are strongly governed by the underlying velocity distribution functions of the constituent species, and can be substantially modified in the presence of superthermal electrons. However, increasing experimental and observational evidence suggests that space plasma particle distributions frequently deviate from thermal equilibrium and exhibit high-energy tails \cite{eawkp15}-\cite{eawkp17}. To account for these deviations, Vasyliunas \cite{eawkp18} introduced the kappa-distribution function as an empirical model to fit satellite observations from OGO-1 and OGO-3 in the terrestrial magnetosphere. The kappa-distribution has since become a standard tool for modeling superthermal electrons in space plasmas, as it successfully captures the enhanced population of energetic particles and significantly modifies the dispersive and nonlinear characteristics of plasma waves \cite{eawkp19}-\cite{eawkp22}. The presence of superthermal electrons is known to strongly influence the existence domain, amplitude, width, and polarity of EA solitary waves, shocks and double layers \cite{eawkp23}-\cite{eawkp26}. In particular, decreasing values of the spectral index $\kappa$ enhance the contribution of energetic electrons, thereby altering the balance between nonlinearity and dispersion. This makes the study of EAWs in $\kappa$-distributed plasmas especially relevant for realistic modeling of space environments such as the Saturn's magnetosphere, solar wind, and astrophysical plasmas \cite{eawkp27}-\cite{eawkp29}.

Another important aspect of nonlinear plasma dynamics is the interaction of nonlinear waves, which is a ubiquitous phenomenon in both laboratory and natural plasma systems \cite{eawkp30}-\cite{eawkp33}.
In addition to standard soliton solutions, special classes of localized excitations such as lump and breather solutions have attracted sustained interest owing to their rich mathematical structure and physical relevance. Lump solitons correspond to rationally localized waveforms that decay algebraically in all spatial directions while retaining pronounced nonlinear features. They are frequently interpreted as a particular manifestation of rogue waves, whose amplitudes can substantially exceed the surrounding background level, sometimes by more than a factor of two.
An oscillatory structure that emerges on or above a soliton background is commonly referred to as a breather. Numerous investigations have reported the occurrence of breather excitations in a variety of fluid and plasma environments, supported by detailed theoretical analyses \cite{eawkp031}-\cite{eawkp035}. Several distinct classes of breather solutions have been identified, most notably the Ma breather and the Akhmediev breather, each characterized by different types of temporal or spatial periodicity.
The concept of lump-type excitations was first introduced by Manakov \textit{et al.} \cite{eawkp036}, opening new avenues for the physical interpretation of strongly localized structures. Since then, both lump and breather solutions have become topics of major importance across diverse research areas, including plasma physics, nonlinear and laser optics, gas dynamics, hydrodynamics, and electromagnetic theory. In the study of soliton interaction, the two-sided Korteweg-de Vries (KdV) equation has been extensively employed to describe resonance phenomena observed in laboratory experiments on shallow water waves, two-core optical fibers, plasma experiments, and fluid-filled elastic tubes \cite{eawkp34}-\cite{eawkp36}. In general, soliton interactions occur through two distinct mechanisms: overtaking collisions, in which solitons propagate in the same direction with different velocities, and head-on collisions (HOC), where solitons move towards each other in opposite directions \cite{eawkp37}. The overtaking collision of solitary waves can be effectively analyzed using the inverse scattering transform technique \cite{eawkp38}. In contrast, HOC require a different theoretical treatment, since they involve the mutual interaction of counter-propagating waves. To investigate HOC, one must consider the evolution of solitary waves traveling in opposite directions, which necessitates the use of appropriate asymptotic expansions to solve the underlying fluid dynamical equations. The standard and widely accepted approach for this purpose is the extended PLK perturbation technique \cite{eawkp39, eawkp40}. The PLK method has been successfully applied to study HOC of solitons in diverse physical systems, including blood flow models \cite{eawkp41}, nonlinear transmission lines \cite{eawkp42}, Bose-Einstein condensates \cite{eawkp43}, and plasma media \cite{eawkp44}-\cite{eawkp48}.
The interaction of nonlinear EA solitary waves (EASWs) constitutes one of the most significant physical processes in space plasmas, as it governs the evolution, stability, and energy transport of localized electrostatic structures frequently observed by spacecraft missions. Several theoretical studies have examined the HOC dynamics of EASWs under different plasma conditions. For instance, Eslami \textit{et al.} \cite{eawkp49} investigated HOC between EA solitons in unmagnetized nonextensive plasmas, demonstrating that such interactions are generally elastic but accompanied by finite phase shifts. Jahangir and Masood \cite{eawkp50} investigated the propagation and interaction of nonlinear EAWs in the presence of superthermal electrons in terrestrial
magnetosphere. More recently, Akter and Hafez \cite{eawkp51} examined the HOC of counter-propagating EA solitons and double layers in unmagnetized, collisionless electron-positron-ion plasmas, revealing the sensitivity of collision outcomes to plasma composition and plasma parameters. Despite these important contributions, existing investigations have been largely restricted to one-dimensional geometry, where transverse perturbations and spatial localization effects are inherently neglected. However, realistic space plasma environments are fundamentally multidimensional. In such settings, two-dimensional effects can introduce qualitatively new physical phenomena that cannot be captured by one-dimensional models. In particular, the transverse balance between nonlinearity and dispersion can lead to the formation of two-dimensional localized structures, such as breathers and lump solitons, which exhibit enhanced spatial localization and distinct stability properties.

The HOC of two-dimensional EA solitons and breathers represents a more realistic interaction scenario, where energy can be redistributed not only along the direction of propagation but also across transverse directions. During such collisions, nonlinear wave coupling may result in phase shifts, transient amplitude modulation, deformation of wave profiles, or localized energy focusing, depending on the plasma parameters and the degree of superthermality. These interaction-induced modifications play a crucial role in regulating energy transfer, particle trapping, and wave-particle interactions in superthermal plasmas commonly encountered in planetary magnetospheres and astrophysical environments. To the best of our knowledge, HOC between lump solitons and single EA solitons in superthermal plasmas have not been explored so far. Lump solitons, which are inherently two-dimensional and algebraically localized in all spatial directions, represent a distinct class of nonlinear excitations that cannot arise in one-dimensional geometries.
Investigating their interaction with conventional EA solitons is therefore essential for understanding multidimensional energy localization, transverse coupling, and nonlinear energy redistribution processes in realistic space plasma environments.
Motivated by these considerations, the present work focuses on the formation and interaction of two-dimensional EA solitons and breathers in superthermal plasmas. By incorporating kappa distributed hot electrons, the study aims to elucidate the combined effects of superthermality and multidimensionality on the nonlinear propagation dynamics of EAWs, with particular emphasis on the HOC processes of solitons, breather-type solutions, and lump solitons, that are directly relevant to observed electrostatic structures in space plasmas. 

\section{Governing Equations}\label{sec2}
Let us consider a three-species unmagnetized plasma having dynamical cold electrons and inertialess non-Maxwellian hot electrons following the kappa distribution and stationary positive ions. To study the nonlinear propagation of EASWs, the governing fluid equations in dimensionless forms are
formulated as \cite{kp3}
\begin{equation}\label{Eq1}
\partial_{t}n_{c}+\partial_{x}(n_{c}v_{x})+\partial_{y}(n_{c}v_{y})=0,
\end{equation}
\begin{equation}\label{Eq2}
\partial_{t}v_{x}+v_{x}\partial_{x}v_{x}+v_{y}\partial_{y}v_{x}-v_{y}^{2}=p\partial_{x}\phi-\frac{p\sigma}{n_{c}}\partial_{x}n_{c},
\end{equation}
\begin{equation}\label{Eq3}
\partial_{t}v_{y}+v_{x}\partial_{x}v_{y}+v_{y}\partial_{y}v_{y}=p\partial_{y}\phi-\frac{p\sigma}{n_{c}}\partial_{y}n_{c},
\end{equation}
\begin{equation}\label{Eq4}
   \partial_{x}^{2}\phi+\partial_{y}^{2}\phi=\left[\frac{1}{p}n_{c}+n_{h}-\left(1+\frac{1}{p}\right)\right],
\end{equation}
where densities of cold and hot electrons are represented by $n_c$ and $n_h$ respectively and normalized by unperturbed densities $n_{c0}$ and $n_{h0}$ . Velocites in $x, y$ directions are represented by $v_x$ and $v_y$ respectively and normalized by the speed $c_e=\sqrt{T_hn_{c0}/m_en_{h0}}$. The electrostatic potential is denoted by $\phi$ and normalized by $T_h/e$. The coordinate of space and time are normalized by the Debye length $\lambda_{de}=\sqrt{T_h/4\pi n_{h0}e^2}$ and inverse cold electron plasma period $\omega_{pc}=\sqrt{4\pi n_{c0}e^2/m_e}$, respectively. Also $\sigma=\frac{T_c}{T_h}$ denotes the temperature ratio of the electron species. The hot electron concentration is represented by $p=\frac{n_{h0}}{n_{c0}}$.
The hot electron density expressed by kappa distribution is written as
\begin{equation}\label{Eq4a}
n_{h}=1+A_{1}\phi+B_{1}\phi^{2}+c_{}\phi^{3}+\dotsc,
\end{equation}
 where $A_1=\frac{\kappa-1/2}{\kappa-3/2}$ and $B_1=\frac{\kappa^2-1/4}{2(\kappa-3/2)^2}$.

\section{Derivation of Two-sided KPE }\label{sec3}
We consider the interaction of two EASWs, denoted by $S_1$ and $S_2$, each characterized by a small but finite amplitude. Initially, the two solitary structures are assumed to be asymptotically well separated and propagate toward each other from opposite directions. As time evolves, the solitons approach, undergo mutual interaction during a finite collision interval, and subsequently separate while continuing their propagation. The amplitudes of both solitary waves are assumed $\sim \epsilon$, where $\epsilon$ is a small, dimensionless perturbation parameter that quantifies the weakness of nonlinearity in the system. The weakly nonlinear nature of the waves implies that the interaction between the two solitary structures is also weak. Consequently, the collision process is expected to be quasi-elastic in nature, meaning that the solitons largely retain their original identities, shapes, and amplitudes after the interaction, while experiencing only small modifications such as phase shifts and slight temporal delays. To investigate the effects of such HOC, we employ the extended PLK perturbation method, which is particularly well suited for analyzing the interaction of counter-propagating solitary waves. Within this approach, multiple stretched space-time scales are introduced to separately describe the slow evolution of each soliton as well as their mutual interaction. Accordingly, the dependent plasma variables are expanded as \cite{kp4, kp5}

\begin{equation}
\begin{aligned}\label{deptrans}
n_{h}&=1+A_{1}\phi+B_{1}\phi^{2}+c_{}\phi^{3}+\dotsc,\\
n_{c}&=1+\epsilon^{2}n_{c_{1}}+\epsilon^{4}n_{c_{2}}+\dotsc,\\
v_{x}&=\epsilon^{2}v_{x_{1}}+\epsilon^{4}v_{x_{2}}+\dotsc,\\
v_{y}&=\epsilon^{3}v_{y_{1}}+\epsilon^{5}v_{y_{2}}+\dotsc,\\
\phi&=\epsilon^{2}\phi^{(1)}+\epsilon^{4}\phi^{(2)}\dotsc .
\end{aligned}
\end{equation}
 Now introducing the stretched independent variable as
\begin{equation}\label{indetrans}
\begin{aligned}
\chi &=\epsilon^2y,\\
\xi &=\epsilon(x-\lambda t)+ \epsilon^2P_0(\eta,\tau)+\epsilon^3P_1(\xi,\eta,\tau)\dotsc,\\
\eta &=\epsilon(x+\lambda t)+ \epsilon^2Q_0(\xi,\tau)+\epsilon^3Q_1(\xi,\eta,\tau)\dotsc,\\
\tau &=\epsilon^3 t.
\end{aligned}
\end{equation}
Let $\xi$ and $\eta$ denote the characteristic coordinates associated with the trajectories of the two counter-propagating solitary waves moving toward each other. The quantities $P_0$ and $Q_0$ which account for the slow phase corrections arising from nonlinear interaction effects, will be determined later in the analysis. We now introduce the following differential operators as
 \begin{equation}\label{operat}
  \begin{aligned}
      \hat{X} &=\partial_\xi+\partial_\eta,\\
      \hat{X'} &=(\partial_\eta P_0) \partial_\xi+(\partial_\xi Q_0) \partial_\eta,\\
       \hat{T} &=-\partial_\xi+\partial_\eta,\\
      \hat{T'} &=(\partial_\eta P_0) \partial_\xi-(\partial_\xi Q_0) \partial_\eta.
  \end{aligned}
 \end{equation}
Hence
\begin{equation}\label{compoperat}
    \begin{aligned}
     \partial_x&=\epsilon\hat{X}+\epsilon^3\hat{X'}\cdots,\\
     \partial_t&= \epsilon\lambda\hat{T} + \epsilon^3\lambda \hat{T'} +\epsilon^3\partial_\tau\cdots,\\
     \partial_y&=\epsilon^2\partial_{\chi}.
    \end{aligned}
\end{equation}
By taking the lower power of $\epsilon$, we obtained
\begin{align}
\lambda\hat{T}n_{c_{1}}&+\hat{X}v_{x_{1}}=0,\\
\lambda\hat{T}v_{x_{1}}&-p\hat{x}\phi^{(1)}+p\sigma\hat{x}n_{c_{1}}=0,\\
\lambda\hat{T}v_{y_{1}}&=p\frac{\partial\phi^{(1)}}{\partial\chi}-p\sigma\frac{\partial n_{c_{1}}}{\partial\chi}=0,\\
\frac{1}{p}n_{c_{1}}&+A_1\phi^{(1)}=0.
\end{align}
    Solving the set of equation, we obtained the phase velocity as
    \begin{align}
        \lambda^2=\frac{1}{A_1}+p\sigma.
    \end{align}
Thus, we derive the relationships between the physical quantities as
\begin{equation}\label{sol1}
\begin{aligned}
   \phi^{(1)} &=\phi_{\xi}^{(1)}+\phi_{\eta}^{(1)} , \hspace{2mm} \phi_{\xi}^{(1)}=\phi_{\xi}^{(1)}(\xi,\chi,\tau),\hspace{2mm} \phi_{\eta}^{(1)}=\phi_{\eta}^{(1)}(\eta,\chi,\tau),\\
n_{c_{1}}&=-pA_1(\phi_{\xi}^{(1)}+\phi_{\eta}^{(1)}),\\
v_{x_{1}}&=-\lambda pA_1(\phi_{\xi}^{(1)}-\phi_{\eta}^{(1)}).
\end{aligned}
\end{equation}
At the subsequent higher order of $\epsilon$, we get
\begin{equation*}
   \begin{aligned}
   &\lambda\hat{T}n_{c_{2}}+\lambda\hat{T}^{\prime}n_{c_{1}}+\partial_{\tau}n_{c_{1}}+\hat{X}v_{x_{2}}+\hat{X}^{\prime}v_{x_{1}}+\hat{X}n_{c_{1}}v_{x_{1}}+\frac{\partial}{\partial\chi}v_{y_{1}}=0,\\
   &\lambda\hat{T}v_{x_{2}}+\lambda\hat{T}^{\prime}v_{x_{1}}+\partial_{\tau}v_{x_{1}}+v_{x_{1}}\hat{X}v_{x_{1}}-p\hat{X}\phi^{(2)}-p\hat{X}^{\prime}\phi^{(1)}+p\sigma\hat{X}n_{c_{2}}+p\sigma\hat{X}^{\prime}n_{c_{1}}-p\sigma n_{c_{1}}\hat{X}n_{c_{1}}=0,\\
   &\lambda\hat{T}v_{y_{2}}+\lambda\hat{T}^{\prime}v_{y_{1}}+\partial_{\tau}v_{y_{1}}+v_{x_{1}}\hat{X}v_{y_{1}}-p\frac{\partial}{\partial\chi}\phi^{(2)}+p\sigma\frac{\partial}{\partial\chi}n_{c_{2}}-p\sigma n_{c_{1}}\frac{\partial n_{c_{1}}}{\partial\chi}=0,\\
   &\frac{1}{p}n_{c_{2}}+A_1\phi^{(2)}=\hat{X}^{2}\phi^{(1)}-\dotsc\{\phi^{(1)}\}^2,\\
   &\frac{\partial v_{y_{1}}}{\partial\xi}=-pA\lambda\frac{\partial\phi_\xi^{(1)}}{\partial\chi},\vspace{8mm}\frac{\partial v_{y_{1}}}{\partial\eta}=pA\lambda\frac{\partial\phi_{\eta}^{(1)}}{\partial\chi}.
   \end{aligned}
\end{equation*}

After calculating the above set of equation we have,
\begin{equation}\label{kp}
\begin{aligned}
4\lambda v_{x_{2}}=\iint\left[\frac{\partial}{\partial\xi}\left(A\frac{\partial\phi_{\xi}^{(1))}}{\partial\tau}-B~\phi_{\xi}^{(1)}\frac{\partial\phi^{(1)}_\xi}{\partial\xi}+C\frac{\partial^{3}\phi_{\xi}^{(1)}}{\partial\xi^{3}}\right)+D\frac{\partial^{2}\phi^{(1)}_{\xi}}{\partial\chi^{2}}\right]d\xi d\eta\\
+\iint\left[\frac{\partial}{\partial\eta}\left(A\frac{\partial\phi_{\eta}^{(1))}}{\partial\tau}-B~\phi_{\eta}^{(1)}\frac{\partial\phi^{(1)}_\eta}{\partial\eta}-C\frac{\partial^{3}\phi_{\eta}^{(1)}}{\partial\eta^{3}}\right)-D\frac{\partial^{2}\phi^{(1)}_{\eta}}{\partial\chi^{2}}\right]d\xi~ d\eta\\\\-\iint(E~\phi^{(1)}_\eta-F\frac{\partial P_{0}}{\partial\eta})\frac{\partial^{2}\phi_{\xi}^{(1)}}{\partial\xi^2}d~\xi~d\eta+\iint(E~\phi^{(1))}_\xi-F\frac{\partial Q_0}{\partial\xi})\frac{\partial^{2}\phi_\eta^{(1)}}{\partial\eta^{2}}d\xi~d\eta,
\end{aligned}
\end{equation}
where
\[
\begin{aligned}
A&=2\lambda A_{1}p,~~~\\
B&=\frac{2B_{1}}{A_{1}}+2~\lambda^{2}A_{1}^{2}p^{2}+p^{2}A_{1},~~~\\
C&=p/A_{1},~~~~\\
D&=\lambda^{2}~pA_{1}~~~~~\\
E&=\frac{2~p~B_{1}}{A_{1}}+p^{2}A_{1}-\lambda^{2}~A^{2}_{1}p^{2},~~~~~\\
F&=4\lambda^{2}A_{1}p.
\end{aligned}
\]
By reassembling the terms of the equation (\ref{kp}) that depend on the stretched variables $\xi$ or $\eta$, together with the slow time variable $\tau$, we obtain the standard KP equations governing the evolution of solitary waves propagating in the right- and left-moving directions, respectively, as follows
\begin{equation}\label{lkp}
    \frac{\partial}{\partial\xi}\left(A\frac{\partial\phi_{\xi}^{(1))}}{\partial\tau}-B~\phi_{\xi}^{(1)}\frac{\partial\phi^{(1)}_\xi}{\partial\xi}+C\frac{\partial^{3}\phi_{\xi}^{(1)}}{\partial\xi^{3}}\right)+D\frac{\partial^{2}\phi^{(1)}_{\xi}}{\partial\chi^{2}}=0,
\end{equation}
\begin{equation}\label{rkp}
    \frac{\partial}{\partial\eta}\left(A\frac{\partial\phi_{\eta}^{(1))}}{\partial\tau}+B~\phi_{\eta}^{(1)}\frac{\partial\phi^{(1)}_\eta}{\partial\eta}-C\frac{\partial^{3}\phi_{\eta}^{(1)}}{\partial\eta^{3}}\right)-D\frac{\partial^{2}\phi^{(1)}_{\eta}}{\partial\chi^{2}}=0.
\end{equation}

The third and fourth terms in equation (\ref{kp}) are not secular in this arrangement, however, they may become secular in a different arrangement. Therefore, we have
\begin{equation}\label{phaseshift12}
 E~\phi^{(1)}_\eta-F\frac{\partial P_{0}}{\partial\eta}=0, \hspace{3mm} E~\phi^{(1))}_\xi-F\frac{\partial Q_0}{\partial\xi}=0.
\end{equation}

\section{Solutions of KPE Using HBM}\label{sec4}
\subsection{Multisoliton and Breather Solution of KPE  (\ref{lkp}) :}

Let $\phi_{\xi}^{(1)} = \Phi(\xi,\chi,\tau)$.
Then equation (\ref{lkp}) becomes
\begin{equation}
\left(A \Phi_{\tau}
- B \Phi \Phi_{\xi}
+ C \Phi_{\xi\xi\xi}\right)_{\xi}
+ D \Phi_{\chi\chi}=0.
\end{equation}

Dividing by $A$, one obtains
\begin{equation}\label{eq:kp-general}
\left(\Phi_{\tau} + \alpha \Phi \Phi_{\xi} - \beta \Phi_{\xi\xi\xi}\right)_{\xi}
+ \gamma \Phi_{\chi\chi}=0,
\end{equation}
where
\[
\alpha=-\frac{B}{A}, \qquad
\beta=-\frac{C}{A}, \qquad
\gamma=\frac{D}{A}.
\]

We now employ the Hirota bilinear method to obtain exact analytical solutions of the KP equation. For this, introduce the transformation \cite{kp3}
\begin{equation}
\Phi = \frac{12 \beta}{\alpha}\frac{\partial^{2}}{\partial\xi^{2}}\left(\ln F(\xi,\chi,\tau)\right).
\end{equation}
This converts Eq.~\eqref{eq:kp-general} into the bilinear form
\begin{equation}\label{eq:bilinear}
\left(D_{\tau}D_{\xi} - \beta D_{\xi}^{4} + \gamma D_{\chi}^{2}\right)
F \cdot F = 0,
\end{equation}
where $D$ is the Hirota bilinear operator.

\subsection*{One-Soliton Solution}
To construct one-soliton solution, consider
\[
F = 1 + e^{\eta_1}, \qquad
\eta_1 = k_1\xi + l_1\chi - \omega_1\tau + n_1.
\]
Substitution of this expression into equation (\ref{eq:bilinear}) yields the dispersion relation
\begin{equation}
\omega_1 = \beta k_1^{3} - \frac{\gamma\, l_1^{2}}{k_1}.
\label{eq:dispersion}
\end{equation}
Thus one-soliton solution of KP Eq. (\ref{eq:kp-general}) is
\begin{equation}\label{s1lkp}
\Phi=\frac{12\beta}{\alpha}k_1^2 \text{sech}^2(\eta_1/2).
\end{equation}
\subsection*{Two-Soliton Solution}

For the generation of the two-soliton solution of the KP equation, we introduce the following form of the auxiliary function:
\begin{equation}
F = 1 + e^{\eta_1} + e^{\eta_2} + A_{12}\, e^{\eta_1+\eta_2},
\label{eq:two-soliton}
\end{equation}
where
\[
\eta_i = k_i\xi + l_i\chi - \omega_i\tau + n_{i}, \qquad i=1,2,
\]
and
\begin{eqnarray}
\omega_i &= \beta k_i^{3} - \frac{\gamma\, l_i^{2}}{k_i},\\
A_{12} &=
\frac{k_1 \omega_2+k_2 \omega_1-\beta(4k_1^3k_2+4k_1k_2^3-6k_1^2k_2^2)+2\gamma l_1l_2}
     {k_1 \omega_2+k_2 \omega_1-\beta(4k_1^3k_2+4k_1k_2^3+6k_1^2k_2^2)+2\gamma l_1l_2}.
\end{eqnarray}

Thus,
\begin{equation}
\Phi(\xi,\chi,\tau)
= \frac{12\beta}{\alpha}\,\frac{\partial^{2}}{\partial\xi^{2}}
\ln\!\Big[\,1 + e^{\eta_1} + e^{\eta_2} + A_{12} e^{\eta_1+\eta_2}\,\Big].
\label{eq:phi-final}
\end{equation}

\subsection*{Breather Solution}

In order to construct the breather solution, we assume that the wave numbers appearing in the two-soliton solution form a pair of complex conjugates, namely \cite{kp5, kp6},
\[
k_1 = m + i n, \quad k_2 = m - i n, \quad
l_1 = p_1 + i q_1, \quad l_2 = p_1 - i q_1.
\]
Choosing $n_2 = n_1=0$ ensures reality of the solution.

Substituting these into equation~(\ref{eq:phi-final}) and simplifying gives
\begin{equation}
\Phi(\xi,\chi,\tau)
= \frac{12\beta}{\alpha}\,\frac{\partial^{2}}{\partial\xi^{2}}
\ln\!\Big[\,1 + 2\,e^{\eta_r}\cos(\eta_i) + A_{12}\,e^{2\eta_r}\,\Big],
\label{eq:breather}
\end{equation}
where
\begin{align}
\eta_r &= m\xi + p_1\chi - \omega_r\tau , \nonumber\\
\eta_i &= n\xi + q_1\chi - \omega_i\tau, \nonumber\\
\omega_r &= \beta(m^3-3mn^2)-\frac{\gamma m(p_1^2-q_1^2)}{m^2+n^2}-\frac{2\gamma p_1q_1n}{m^2+n^2} ,\nonumber\\
\omega_i &= \beta(3m^2n-n^3)+\frac{\gamma n(p_1^2-q_1^2)}{m^2+n^2}-\frac{2\gamma mp_1q_1}{m^2+n^2} ,\nonumber\\
A_{12} &= \frac{4n\omega_i+16\beta n^4+4\gamma q_1^2}{4m\omega_r-16\beta m^4+4\gamma p_1^2}.
\end{align}

\subsection{Multisoliton, Breather and Lump Solution of KPE  (\ref{rkp}):}
Let $\phi_{\eta}^{(1)} = \Theta(\eta,\chi,\tau)$.
Then equation (\ref{rkp}) becomes
\begin{equation}
\left(A \Theta_{\tau}
+ B \Theta \Theta_{\eta}
- C \Theta_{\eta\eta\eta}\right)_{\eta}
- D \Theta_{\chi\chi}=0.
\end{equation}

Dividing by $A$, one obtains
\begin{equation}\label{eq:rkp-general}
\left(\Theta_{\tau} + \alpha  \Theta \Theta_{\eta} + \beta \Theta_{\eta\eta\eta}\right)_{\eta}
- \gamma \Theta_{\chi\chi}=0,
\end{equation}
where
\[
\alpha=-\frac{B}{A}, \qquad
\beta=-\frac{C}{A}, \qquad
\gamma=\frac{D}{A}.
\]

Following the same procedure employed for Eq. (\ref{eq:kp-general}), we introduce the transformation
\begin{equation}
 \Theta = \frac{12 \beta}{\alpha}\frac{\partial^{2}}{\partial\eta^{2}}\left(\ln G(\eta,\chi,\tau)\right),
\end{equation}
into Eq.~\eqref{eq:rkp-general}, thereby obtaining its corresponding Hirota bilinear form, as given below:
\begin{equation}\label{req:bilinear}
\left(D_{\tau}D_{\eta} + \beta D_{\eta}^{4} - \gamma D_{\chi}^{2}\right)
G \cdot G = 0,
\end{equation}
where $D$ is the Hirota bilinear operator.

\subsection*{One-Soliton Solution}
For the construction of the one-soliton solution, we define the auxiliary functions as follows:

\[
G = 1 + e^{\theta_1}, \qquad
\theta_1 = K_1\eta + L_1\chi - \Omega_1\tau + N_1.
\]
Substitution of this expression into equation (\ref{req:bilinear}) yields the dispersion relation
\begin{equation}
\Omega_1 = -\beta K_1^{3} + \frac{\gamma\, L_1^{2}}{K_1}.
\label{eq:rdispersion}
\end{equation}
Thus, one-soliton solution of KPE (\ref{rkp}) is given by
\begin{equation}\label{s1rkp}
\Theta=\frac{12\beta}{\alpha}K_1^2 \text{sech}^2(\theta_1/2)
\end{equation}

\subsection*{Two-Soliton Solution}

For two-soliton solution, we consider
\begin{equation}
G = 1 + e^{\theta_1}+ e^{\theta_2} + B_{12}\, e^{\theta_1+\theta_2},
\label{req:two-soliton}
\end{equation}
where
\[
\theta_1 = K_i\eta + L_i\chi - \Omega_i\tau + N_i., \qquad i=1,2,
\]
and
\begin{eqnarray}
\Omega_i &= -\beta K_i^{3} + \frac{\gamma\, L_i^{2}}{K_i},\\
B_{12} &=
\frac{K_1 \Omega_2+K_2 \Omega_1+\beta(4K_1^3K_2+4K_1K_2^3-6K_1^2K_2^2)-2\gamma L_1L_2}
     {K_1 \Omega_2+K_2 \Omega_1+\beta(4K_1^3K_2+4K_1K_2^3+6K_1^2K_2^2)-2\gamma L_1L_2}.
\end{eqnarray}

Thus, the two-soliton solution of Eq. (\ref{eq:rkp-general}) is given by
\begin{equation}
\Theta(\eta,\chi,\tau)
= \frac{12\beta}{\alpha}\,\frac{\partial^{2}}{\partial\eta^{2}}
\ln\!\Big[\,1 + e^{\theta_1} + e^{\theta_2} + B_{12} e^{\theta_1+\theta_2}\,\Big].
\label{req:phi-final}
\end{equation}

\subsection*{Breather Solution}
The breather solution can be constructed from the two-soliton solution by assuming that the associated wave numbers form a pair of complex conjugates \cite{kp5,kp6}, which leads to a localized, oscillatory wave structure. We consider

\[
K_1 = M + i N, \quad K_2 = M - i N, \quad
L_1 = P_1 + i Q_1, \quad L_2 = P_1 - i Q_1.
\]
Choosing $N_2 = N_1=0$ ensures reality of the solution.

Substituting these into equation~(\ref{req:phi-final}) and simplifying, the solution is obtained as
\begin{equation}
\Theta(\eta,\chi,\tau)
= \frac{12\beta}{\alpha}\,\frac{\partial^{2}}{\partial\eta^{2}}
\ln\!\Big[\,1 + 2\,e^{\theta_r}\cos(\theta_i) + B_{12}\,e^{2\theta_r}\,\Big],
\label{req:breather}
\end{equation}
where
\begin{align}
\theta_r &= M\eta + P_1\chi - \Omega_r\tau , \nonumber\\
\eta_i &= N\eta + Q_1\chi - \Omega_i\tau, \nonumber\\
\Omega_r &= -\beta(M^3-3MN^2)+\frac{\gamma M(P_1^2-Q_1^2)}{M^2+N^2}+\frac{2\gamma P_1Q_1N}{M^2+N^2} ,\nonumber\\
\Omega_i &= -\beta(3M^2N-N^3)-\frac{\gamma N(P_1^2-Q_1^2)}{m^2+N^2}+\frac{2\gamma MP_1Q_1}{M^2+N^2} ,\nonumber\\
B_{12} &= \frac{4N\Omega_i-16\beta N^4-4\gamma Q_1^2}{4M\Omega_r+16\beta M^4-4\gamma P_1^2}.
\end{align}

\subsection*{Lump Solution}
For lump solution of KPE (\ref{eq:rkp-general}), we consider the dependent variable $G(\eta,\chi,\tau)$ as follows \cite{kp7}
\begin{equation}\label{lumps}
G=f^2+g^2+b_9, \hspace{2mm} f=b_1\eta+b_2\chi+b_3\tau+b_4, \hspace{2mm} g=b_5\eta+b_6\chi+b_7\tau+b_8,
\end{equation}
where $b_i, 1\leq i\leq 9$, are the real parameter.
Using the value of $G$, we have,
\begin{equation}
\begin{cases}
b_1=b_1,\hspace{2mm}b_2=b_2,\hspace{2mm}b_3=\frac{\gamma(b_1b_2^2-b_1b_6^2+2b_2b_5b_6)}{b_1^2+b_5^2},\\
b_4=b_4,\hspace{2mm}b_5=b_5,\hspace{2mm}b_6=b_6,\hspace{2mm}b_7=\frac{\gamma(2b_1b_2b_6-b_5b_2^2-b_5b_6^2)}{b_1^2+b_5^2},\\
b_8=b_8, \hspace{2mm} b_9=\frac{3\beta}{\gamma}\frac{(b_1^2+b_5^2)^3}{(b_1b_6-b_2b_5)^2}.
\end{cases}
\end{equation}
with the condition \begin{equation}
\Delta:=b_1b_6-b_2b_5\neq 0
\end{equation}
Therefore, lump solution of the KPE (\ref{eq:rkp-general}) is of the form
\begin{equation}\label{slump}
\Theta(\eta,\chi,\tau)=\frac{12\beta}{\alpha}\frac{\left(2(b_1^2+b_5^2)G-4(b_1f+b_5g)\right)}{G^2}
\end{equation}

\begin{figure}[ht]
	\centering
	\subfigure[]{\includegraphics[width=0.31\linewidth]{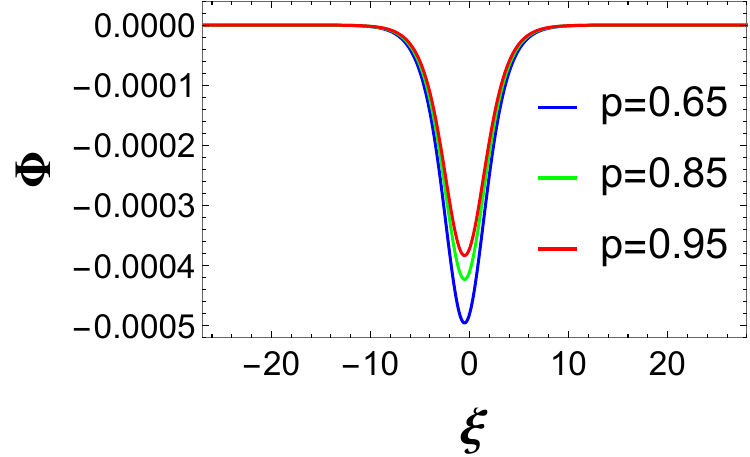}\label{f:1a}}
	\hfill
	\subfigure[]{\includegraphics[width=0.31\linewidth]{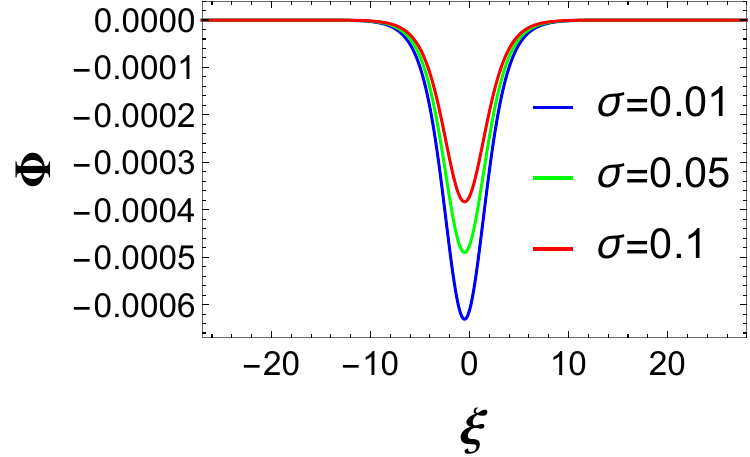}\label{f:1b}}
    \hfill
	\subfigure[]{\includegraphics[width=0.31\linewidth]{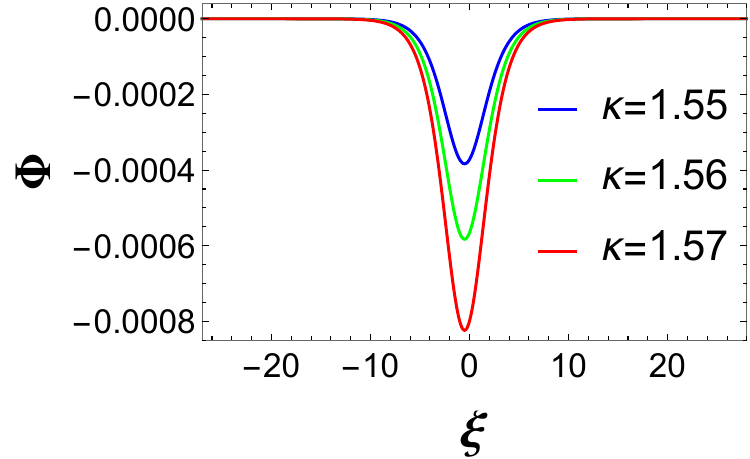}\label{f:1c}}
		
	\caption{Effect of various plasma parameters on the one-soliton solution (\ref{s1lkp}) of KPE (\ref{eq:kp-general}) taking $k_1=0.7$, $l_1=2.5$, $n_1=0.1$, $\tau=0$ and $\chi=0.1$: (a) the impact of hot electron concentration $p$, (b) the impact of the cold-to-hot electron temperature ratio $\sigma$ and (c) the impact of the spectral index $\kappa$.}\label{f:1}
\end{figure}

\section{Result And Discussion}\label{sec5}
We now investigate the nonlinear propagation characteristics of EASWs using plasma parameters inferred from Cassini spacecraft observations in Saturn's magnetosphere. The Cassini mission has provided comprehensive measurements of electron densities, temperature ratios, and the presence of superthermal electron populations in different regions of the Saturnian plasma environment. The plasma parameters characterizing this region are given by \cite{kp8} $n_{h0} \sim  0.18 cm^{-3}$, $n_{c0} \sim  0.21 cm^{-3}$, $T_h \sim 1000 eV $, and $T_c \sim 10.2 eV $. By varying the relevant parameters within observationally realistic ranges, we examine the formation and interaction of EA multisolitons, breathers and lump structures.

\begin{figure}[ht]
	\centering
	\subfigure[]{\includegraphics[width=0.31\linewidth]{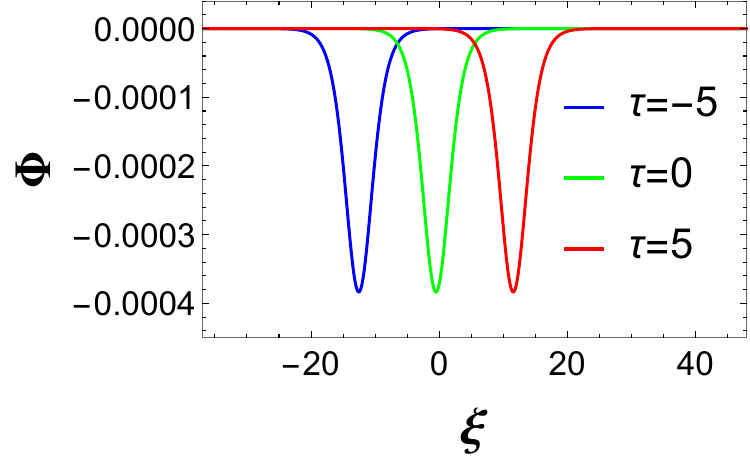}\label{f:2a}}
	\hfill
	\subfigure[]{\includegraphics[width=0.31\linewidth]{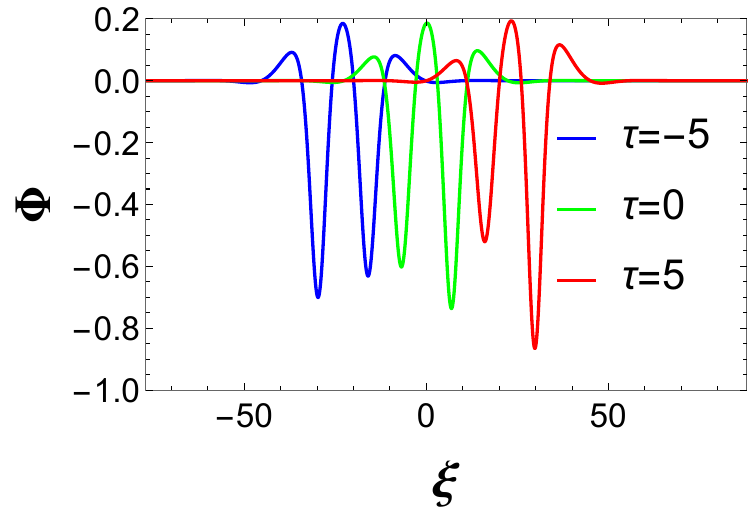}\label{f:2b}}
    \hfill
	\subfigure[]{\includegraphics[width=0.31\linewidth]{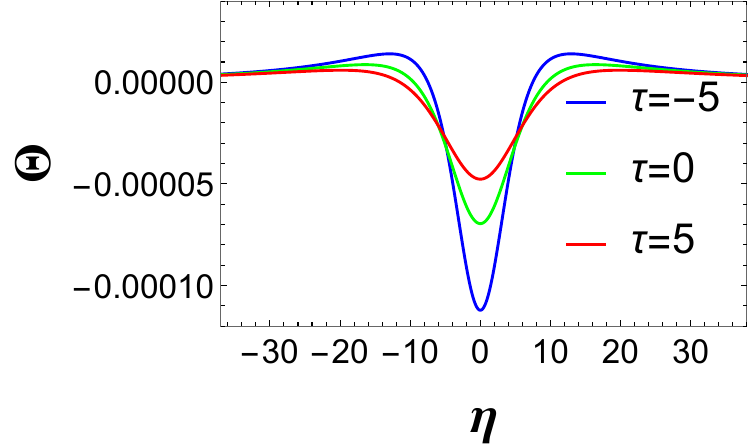}\label{f:2c}}
	\hfill
	\subfigure[]{\includegraphics[width=0.31\linewidth]{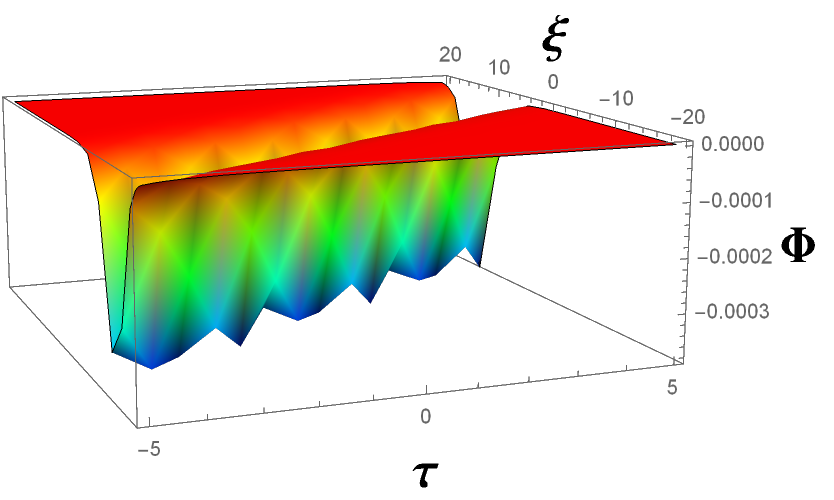}\label{f:2d}}
	\hfill
	\subfigure[]{\includegraphics[width=0.31\linewidth]{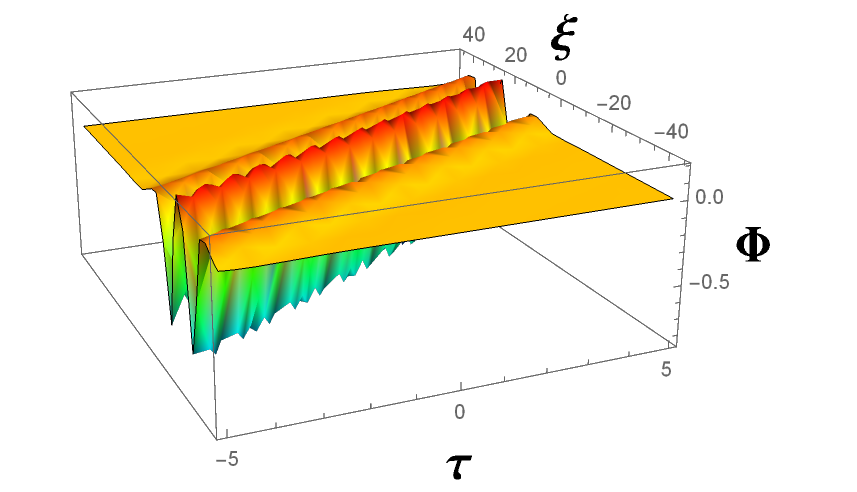}\label{f:2e}}
	\hfill
	\subfigure[]{\includegraphics[width=0.31\linewidth]{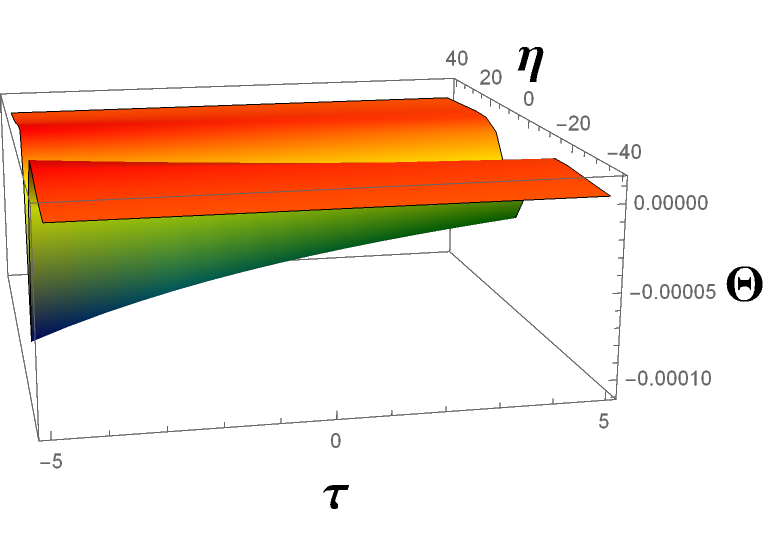}\label{f:2f}}	
	\caption{Temporal evolution on one-soliton solution (\ref{s1lkp}), breather solution (\ref{eq:breather}) and lump solution (\ref{slump}), taking $k_1=0.7$, $l_1=2.5$, $n_1=0.1$, $m=0.23$, $n=0.29$, $p_1=0.36$, $q_1=0.35$, $b_1=0.1$, $b_2=0.1$, $b_4=0$, $b_5=0$, $b_6=0.1$, $b_8=1$, $\chi=0.1$, $\kappa=1.55$, $\sigma=0.1$ and $p=0.95$: here (a), (b), (c) are the 2D profile of one-soliton solution , breather solution, lump solution and (d), (e), (f) are their corresponding 3D profile, respectively. }\label{f:2}
\end{figure}

\begin{figure}[ht]
	\centering
	\subfigure[]{\includegraphics[width=0.31\linewidth]{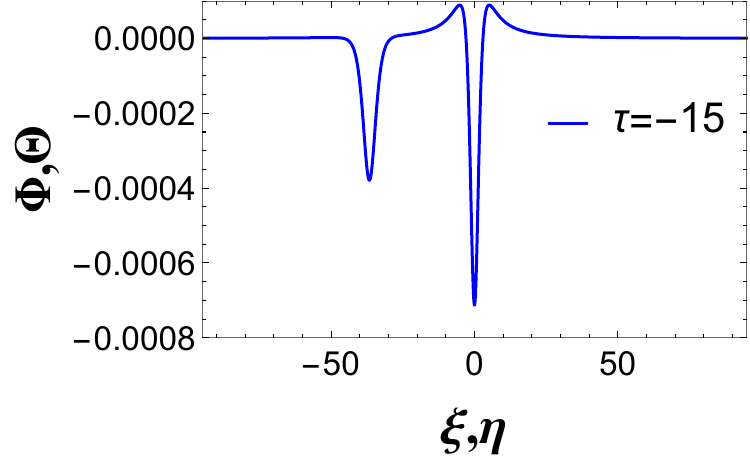}\label{f:3a}}
	\hfill
	\subfigure[]{\includegraphics[width=0.31\linewidth]{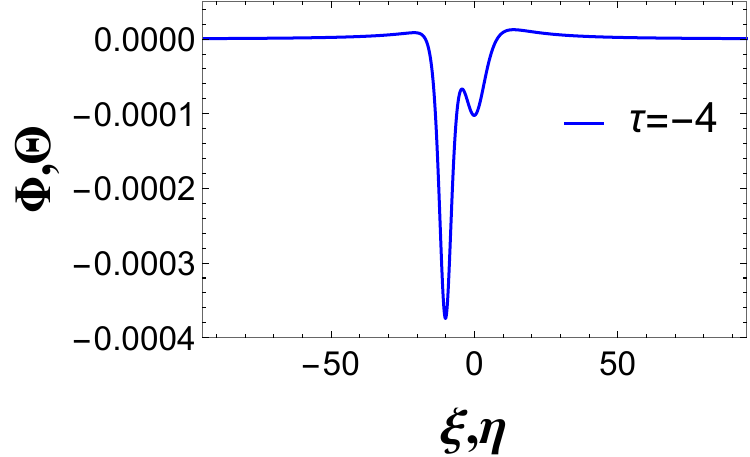}\label{f:3b}}
    \hfill
	\subfigure[]{\includegraphics[width=0.31\linewidth]{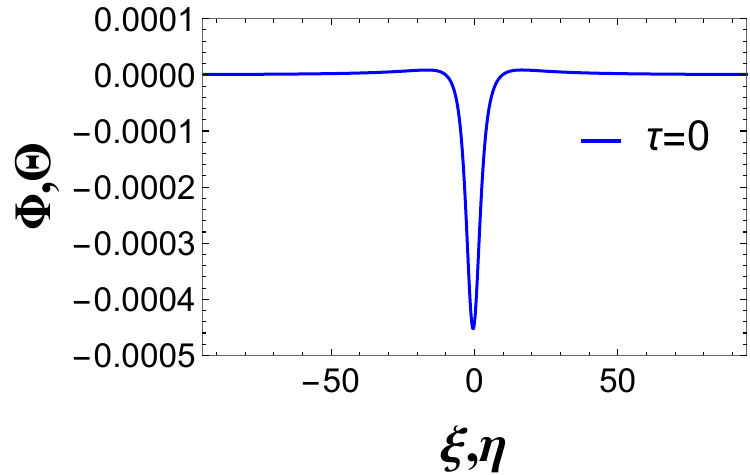}\label{f:3c}}
	\hfill
	\subfigure[]{\includegraphics[width=0.31\linewidth]{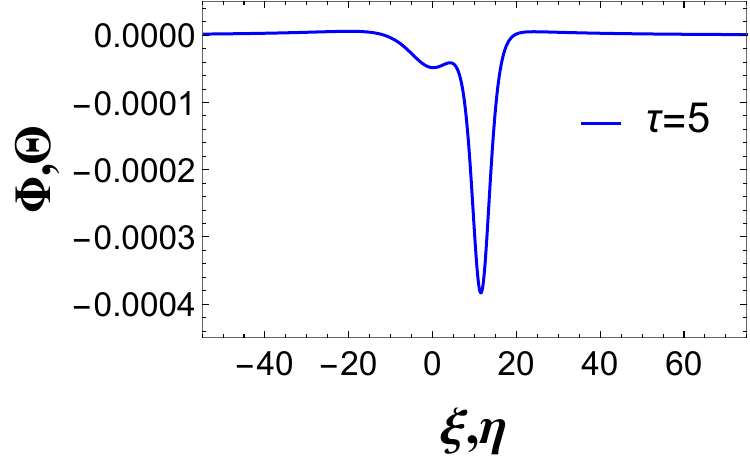}\label{f:3d}}
	\hfill
	\subfigure[]{\includegraphics[width=0.31\linewidth]{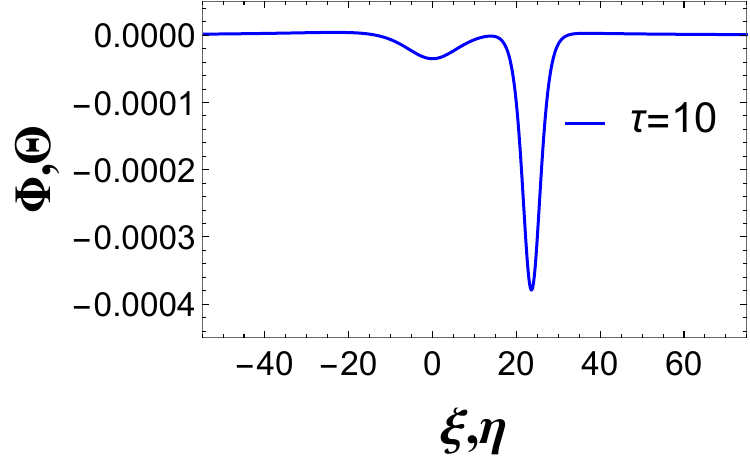}\label{f:3e}}
	\hfill
	\subfigure[]{\includegraphics[width=0.36\linewidth]{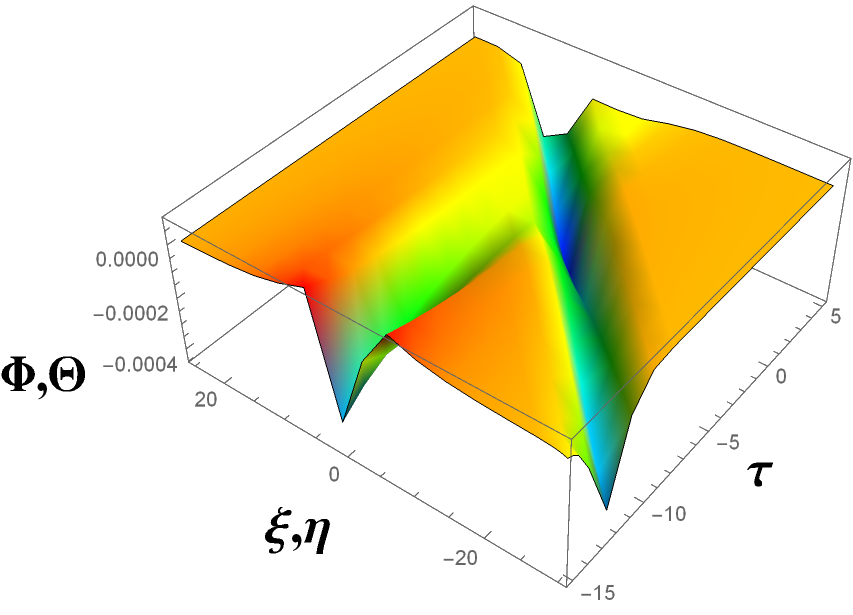}\label{f:3f}}	
	\caption{HOC of one-soliton profile (\ref{s1lkp}) with lump profile (\ref{slump}) for different values of $\tau$ with $k_1=0.7$, $l_1=2.5$, $n_1=0.1$, $b_1=0.1$, $b_2=0.1$, $b_4=0$, $b_5=0$, $b_6=0.1$, $b_8=1$, $\chi=0.1$, $\kappa=1.55$, $\sigma=0.1$ and $p=0.95$. }\label{f:3}
\end{figure}

Figures 1(a)-1(c) illustrate the influence of key plasma parameters on the profile of the one-soliton solution (24) of the KP equation (20). Specifically, Fig. 1(a) shows the effect of the hot electron concentration $p$, Fig. 1(b) depicts the role of the cold-to-hot electron temperature ratio $\sigma$ and Fig. 1(c) demonstrates the impact of the spectral index $\kappa$. It is evident from Fig. 1(a) that an increase in the hot electron concentration leads to a reduction in the amplitude of the EASW. Physically, a larger population of hot electrons enhances charge screening and weakens the nonlinear steepening of the wave, thereby reducing the soliton amplitude. From Fig. 1(b), it is observed that the amplitude of the one-soliton decreases with increasing values of the temperature ratio $\sigma$. Since a higher $\sigma$ corresponds to a reduction in the effective hot electron temperature, the restoring force provided by the hot electron pressure is weakened. This reduction in restoring force resulting in a lower-amplitude solitary structure. Figure 1(c) highlights the role of superthermal electrons by varying the spectral index $\kappa$. As $\kappa$ increases, the electron velocity distribution approaches the Maxwellian limit, reducing the contribution of high-energy superthermal electrons. Consequently, the nonlinearity in the system is enhanced, leading to an increase in the amplitude of the EA one-soliton. This result underscores the strong sensitivity of EASW characteristics to the degree of superthermality present in the plasma. Figures 2(a), 2(b) and 2(c) illustrate the temporal evolution of the EA one-soliton solution (\ref{s1lkp}), the breather solution (\ref{eq:breather}), and the lump solution (\ref{slump}), respectively. These panels demonstrate the distinct dynamical behaviors of the nonlinear EA structures supported by the present plasma model. The one-soliton maintains a stable, localized profile during propagation, reflecting a robust balance between nonlinearity and dispersion. In contrast, the breather solution exhibits a localized structure with periodic temporal modulation of its amplitude, indicating an oscillatory exchange of energy within the wave packet. In figure 2(c) the lump solution, represents a rationally localized excitation with energy concentrated within a short temporal interval. Unlike the shape-preserving soliton and the periodically modulated breather, the lump exhibits a strong but brief amplification near a specific time. Away from this point, the structure broadens and its amplitude decreases, indicating a return of energy to the background, while the coherent nature of the waveform remains intact. Accordingly the lump structure is vanishes after a certain time. The corresponding 3D profiles of these nonlinear structures are presented in Figs. 2(d)-2(f). These 3D representations clearly reveal the spatial localization and amplitude distribution of the soliton, breather, and lump structures, respectively. Figures 3(a)-3(e) illustrate the spatiotemporal evolution of HOC between an EA one-soliton and lump structure at different times. Prior to the interaction, the two nonlinear excitations propagate toward each other while preserving their individual identities and localized profiles, indicating their stability in the weakly nonlinear regime. As the waves approach, nonlinear coupling between the soliton and the localized lump becomes increasingly significant. At the collision stage ($\tau=0$), shown in Fig. 3(c), the one-soliton and the lump structure completely overlap in space, forming a transient composite structure with enhanced localization. This temporary merging reflects a strong but short-lived nonlinear interaction, during which energy is redistributed between the interacting modes without leading to wave breaking or structural collapse. Following the interaction, the two nonlinear structures separate and continue to propagate in opposite directions, as shown in Figs. 3(d) and 3(e). Figure 3(f) presents the 3D profile of the interacting one-soliton and lump structures, providing a clear visualization of their spatial overlap and post-collision separation. The re-emergence of the nonlinear structures with nearly unchanged amplitudes and profiles after the interaction signifies that the HOC proceeds in a quasi-elastic manner.

\begin{figure}[ht]
	\centering
	\subfigure[]{\includegraphics[width=0.31\linewidth]{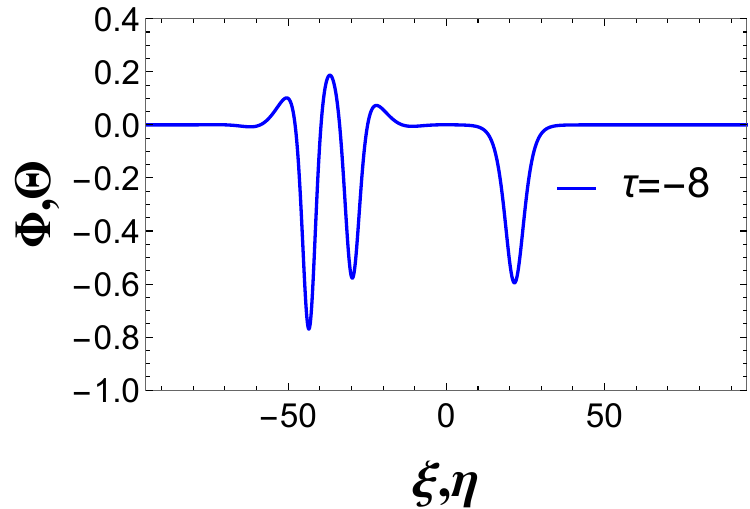}\label{f:4a}}
	\hfill
	\subfigure[]{\includegraphics[width=0.31\linewidth]{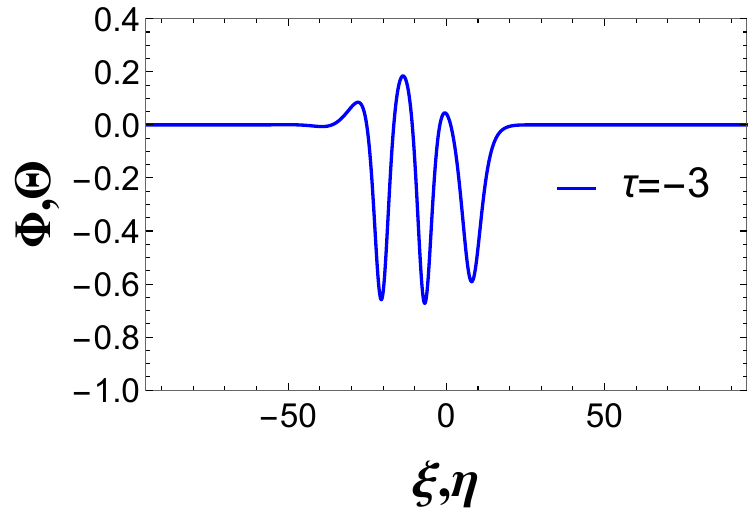}\label{f:4b}}
    \hfill
	\subfigure[]{\includegraphics[width=0.31\linewidth]{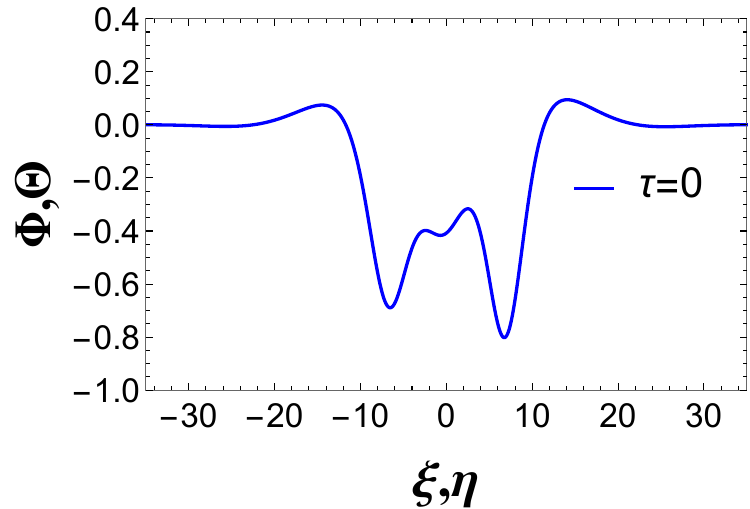}\label{f:4c}}
	\hfill
	\subfigure[]{\includegraphics[width=0.31\linewidth]{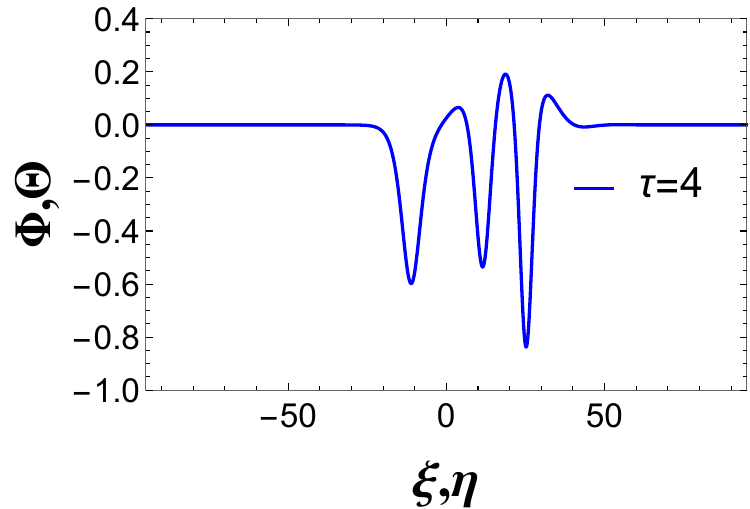}\label{f:4d}}
	\hfill
	\subfigure[]{\includegraphics[width=0.31\linewidth]{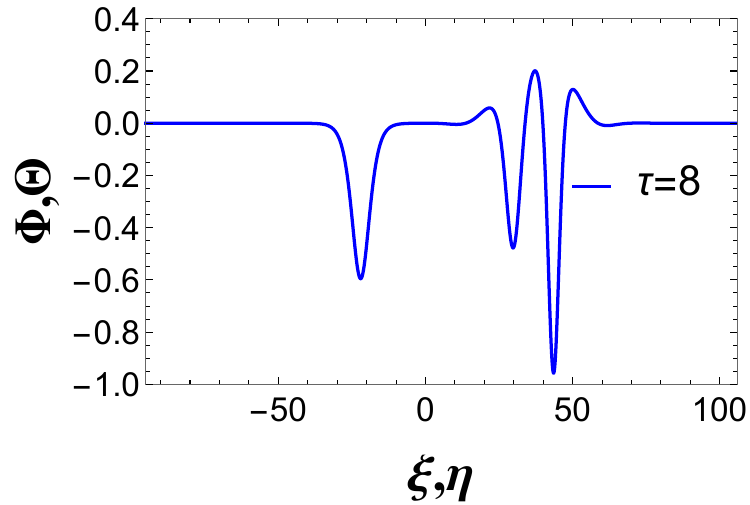}\label{f:4e}}
	\hfill
	\subfigure[]{\includegraphics[width=0.35\linewidth]{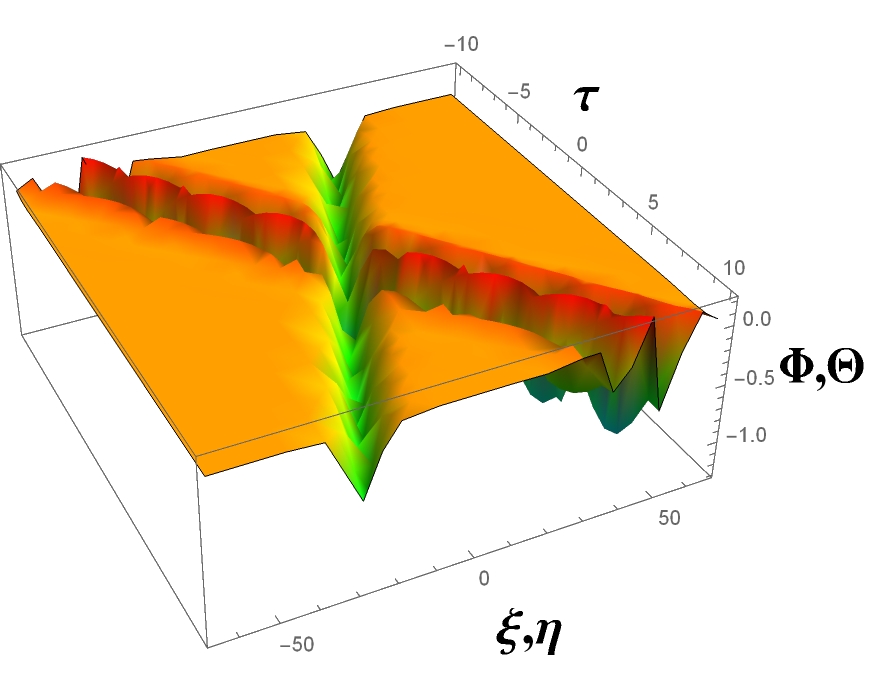}\label{f:4f}}	
	\caption{HOC of breather profile (\ref{eq:breather}) with one-soliton profile (\ref{s1rkp}) for different values of $\tau$ with $m=0.23$, $n=0.29$, $p_1=0.36$, $q_1=0.35$, $K_1=0.5$, $L_1=0.2$, $N_1=0.1$, $\chi=0.1$, $\kappa=1.55$, $\sigma=0.1$ and $p=0.95$. }\label{f:4}
\end{figure}

\begin{figure}[ht]
	\centering
	\subfigure[]{\includegraphics[width=0.31\linewidth]{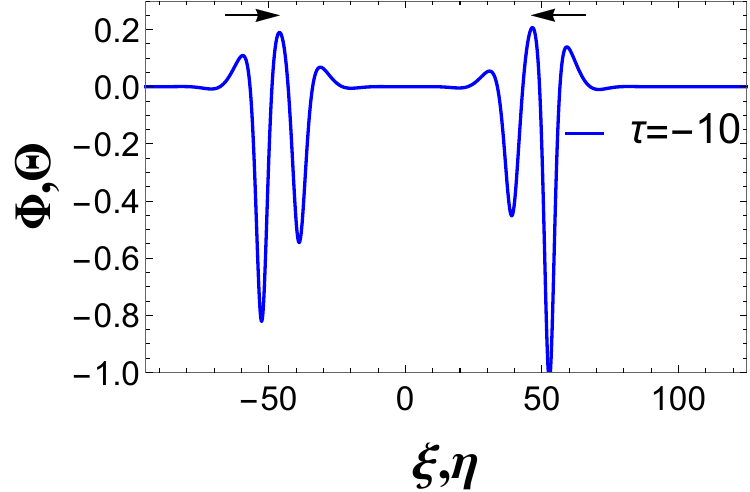}\label{f:5a}}
	\hfill
	\subfigure[]{\includegraphics[width=0.31\linewidth]{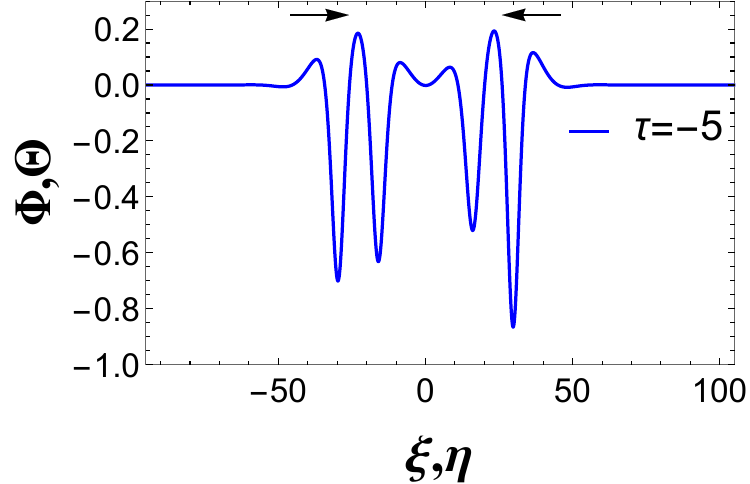}\label{f:5b}}
    \hfill
	\subfigure[]{\includegraphics[width=0.31\linewidth]{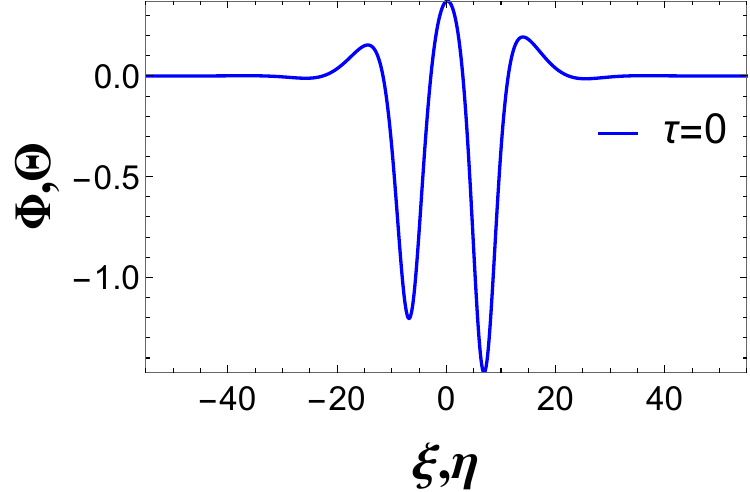}\label{f:5c}}
	\hfill
	\subfigure[]{\includegraphics[width=0.31\linewidth]{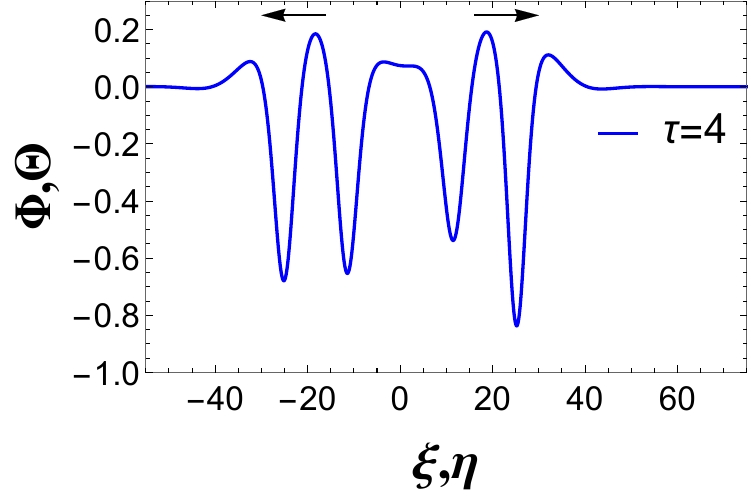}\label{f:5d}}
	\hfill
	\subfigure[]{\includegraphics[width=0.31\linewidth]{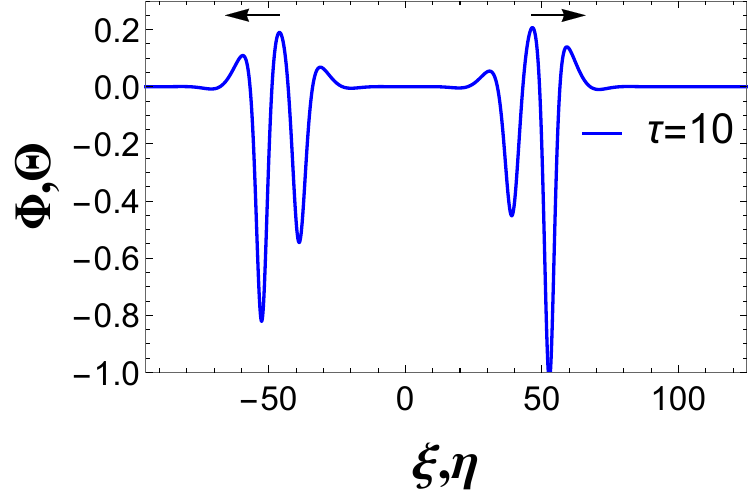}\label{f:5e}}
	\hfill
	\subfigure[]{\includegraphics[width=0.35\linewidth]{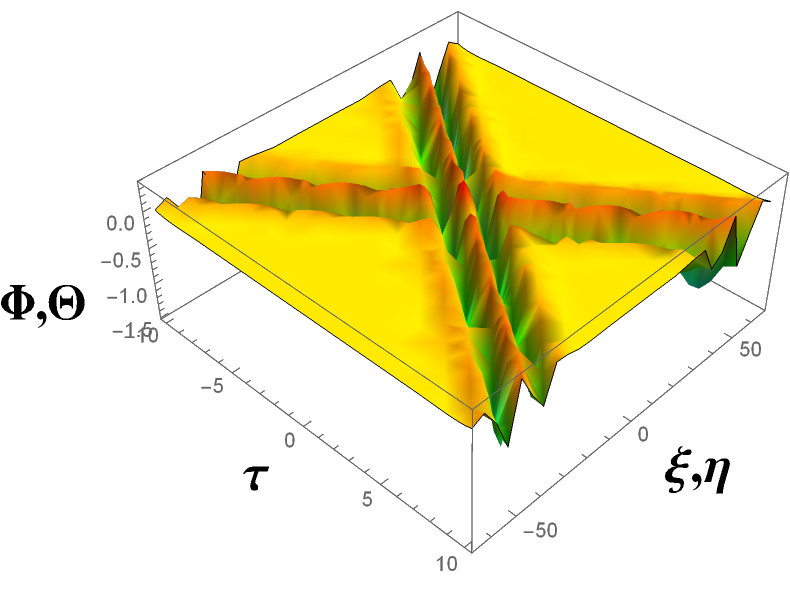}\label{f:5f}}	
	\caption{HOC of breather profile (\ref{eq:breather}) with another breather profile (\ref{req:breather}) for different values of $\tau$ with $m=0.23=M$, $n=0.29=N$, $p_1=0.36=P_1$, $q_1=0.35=Q_1$, $\chi=0.1$, $\kappa=1.55$, $\sigma=0.1$ and $p=0.95$. }\label{f:5}
\end{figure}

Figures 4(a)-4(e) depict the temporal evolution of HOC between EA breather [Eq. (\ref{eq:breather})] and one-soliton structure [Eq. (\ref{s1rkp})]. In the pre-collision stage, both nonlinear modes propagate toward each other without distortion: the breather is characterized by a spatially localized oscillatory envelope, while the soliton behaves as a coherent, non-oscillatory pulse. As the separation between the two structures decreases, their mutual interaction becomes governed by nonlinear mode coupling, allowing the breather's internal oscillations to interact with the soliton's localized electric field.
During the collision, the breather and soliton form a transient compound structure in which the breather oscillations are temporarily compressed and amplified by the soliton-induced localization. This interaction facilitates short-lived energy concentration in the collision region, rather than irreversible wave breaking or dispersion. Following the interaction, the two excitations decouple and propagate away from each other, retaining their characteristic profiles with only minor phase adjustments.
Figures 5(a)-5(e) present the head-on interaction between two counter-propagating EA breather structures given by Eqs. (\ref{eq:breather}) and (\ref{req:breather}). Before the collision, each breather propagates independently, maintaining its localized oscillatory envelope and internal modulation. As the breathers approach one another, their interaction is governed by nonlinear coupling between their oscillatory modes. At the collision instant $\tau=0$, Fig. 5(c) shows a complete spatial overlap of the counter-propagating breathers, forming a highly localized composite structure with enhanced oscillatory amplitude. This stage corresponds to a temporary concentration of wave energy in the interaction region, resulting from constructive interference of the breather envelopes and their internal carrier oscillations.
Following the interaction, the breathers separate and re-emerge with nearly unchanged amplitudes and oscillatory patterns, as revealed by the near mirror symmetry between Figs. 5(a) and 5(e), apart from the reversal in propagation direction. This behavior confirms the quasi-elastic nature of breather-breather collisions, where energy remains largely confined within the oscillatory envelopes and is redistributed primarily through phase adjustments.

\section{Conclusion}\label{sec6}

In this work, the nonlinear dynamics of two dimensional EASWs and breathers in superthermal plasmas have been systematically explored. Starting from a plasma model composed of cold inertial electrons, superthermal hot electrons, and stationary ions, the extended PLK method was employed to derive a pair of two-sided KP equations describing counter-propagating EA modes. This formulation allows a self-consistent treatment of multidimensional effects and HOC processes, which are inherently absent in one dimensional models. Exact analytical solutions in the form of solitons, breathers, and lump structures were obtained using the Hirota bilinear technique. The parametric analysis demonstrates that the amplitude and localization of EA solitary structures are strongly influenced by the hot electron concentration, the cold-to-hot electron temperature ratio, and the superthermality index. Enhanced superthermal effects were found to significantly modify the restoring force and nonlinearity, thereby altering the wave profiles. The interaction analysis reveals that soliton-soliton collisions are predominantly quasi-elastic, characterized by transient overlap and finite phase shifts with minimal energy exchange. In contrast, breather-soliton interactions involve coupling between coherent solitary pulses and internal oscillatory modes, leading to temporary amplitude modulation and asymmetric energy redistribution. Breather-breather collisions exhibit symmetric mode coupling, producing stronger but short lived energy localization, followed by balanced post collision recovery of the interacting structures. These differences underline distinct nonlinear energy redistribution mechanisms associated with oscillatory and non oscillatory EA modes. The inclusion of two dimensional effects is essential for describing realistic interaction dynamics in space plasma environments, particularly in Saturn's magnetospheric environments, where localized electrostatic structures and broadband electrostatic fluctuations are commonly observed.

\section*{Author Contribution}
\textbf{Prasanta Chatterjee:} Conceptualization, supervision, and critical revision of the manuscript. \textbf{Jayshree Mondal:} Analytical calculations, figure preparation, contribution to writing specific sections of the manuscript, and review and editing. \textbf{Biswajit Sahu:} Preparation of the original draft and review and editing of the manuscript. This paper has been read and approved by all authors.

\section*{Declaration of competing interest} The authors declare that they have no competing interests.
\section*{Acknowledgements}
Jayshree Mondal thanks the University Grants Commission (UGC), India, for providing
financial support under the Junior Research Fellowship Program (NTA Ref. No.
-221610016975).
\section*{Funding}
This research work did not receive any external funding.
\section*{Data Availability} The authors confirm that all data supporting the results of this study are included in the manuscript.


\newpage

\begin{thebibliography}{70}
\bibitem{eawkp1} D. Henry and J. P. Treguier, J. Plasma Phys. \textbf{8}, 311 (1972).
\bibitem{eawkp3} S. Ikezawa and Y. Nakamura, J. Phys. Soc. Japan \textbf{50}, 962 (1981).
\bibitem{eawkp4} S. Chowdhury, S. Biswas, N. Chakrabarti, and R. Pal, Phys. Plasmas \textbf{24}, 062111 (2017).
\bibitem{eawkp5} R. L. Tokar and S. P. Gary, Geophys. Res. Lett. \textbf{11}, 1180 (1984).
\bibitem{eawkp8} N. Dubouloz, R. Pottelette, M. Malingre, and R. Treumann, Geophys. Res. Lett. \textbf{18}, 155 (1991).
\bibitem{eawkp9} F. S. Mozer, R. Ergun, M. Temerin, C. Cattell, J. Dombeck, and J. Wygant, Phys. Rev. Lett. \textbf{79}, 1281 (1997).
\bibitem{eawkp10} G. T. Delory \textit{et al.}, Geophys. Res. Lett. \textbf{25}, 2069 (1998).
\bibitem{eawkp12} C. A. Cattell \textit{et al.}, Geophys. Res. Lett. \textbf{26}, 425 (1999).
\bibitem{eawkp13} E. C. J. Sittler, K. W. Ogilvie, and J. D. Scudder, J. Geophys. Res., \textbf{88}(A11), 8847 (1983).
\bibitem{eawkp14} D. T. Young, \textit{et al.}, Science \textbf{307}, 1262 (2005).
\bibitem{eawkp15} R. A. Cairns, A. A. Mamun, R. Bingham, and R. Bostr\"om, Geophys. Res. Lett. \textbf{22}, 2709 (1995).
\bibitem{eawkp16} Y. Futaana, S. Machida, Y. Saito, A. Matsuoka, and H. Hayakawa, J. Geophys. Res. \textbf{108}, 1025 (2003).
\bibitem{eawkp17} G. Livadiotis, \textit{Kappa Distributions: Theory and Applications in Plasmas} (Amsterdam: Elsevier, 2017).
\bibitem{eawkp18} V. M. Vasyliunas, J. Geophys. Res. \textbf{73}, 2839 (1968).
\bibitem{eawkp19} M. A. Hellberg and R. L. Mace, Phys. Plasmas \textbf{9}, 1495 (2002).
\bibitem{eawkp20} C.-R. Choi, K.-W. Min, and T.-N. Rhee, Phys. Plasmas \textbf{18}, 092901 (2011).
\bibitem{eawkp21} T. Akter, F. Deeba, and K.-A. Hassan,  IEEE Trans. Plasma Sci. \textbf{44}, 1449 (2016).
\bibitem{eawkp22} M. Mehdipoor, Eur. Phys. J. Plus \textbf{135}, 299 (2020).
\bibitem{eawkp23} H. Chen and S. Q. Liu, Astrophys. Space Sci. \textbf{339}, 179 (2012).
\bibitem{eawkp24} C. Bedi, and C. B. Singh, Astrophys. Space Sci. \textbf{344}, 161 (2013).
\bibitem{eawkp25} D. Dutta and K. S. Goswami, J. Plasma Phys. \textbf{85}, 905850308 (2019).
\bibitem{eawkp26} A. Atteya, A. Saha, P. K. Karmakar, and E. M. El-Bayoumi, Waves in Random and Complex Media, (2024), https://doi.org/10.1080/17455030.2024.2358113.
\bibitem{eawkp27} S. Devanandhan, S.V. Singh, G.S. Lakhina, and R. Bharuthram, Commun. Nonlinear Sci. Num. Simul. \textbf{22}, 1322 (2015).
\bibitem{eawkp28} R. Jahangir and W. Masood, Phys. Plasmas \textbf{27}, 042105 (2020).
\bibitem{eawkp29} I. E. Ibrahim, H. I. A. Gawad, M. Al-Dossari, and N. S. A. EL-Gawaad, Phys. Fluids \textbf{36}, 107123 (2024).

\bibitem{eawkp30} R. C. Davidson, \textit{Methods in Nonlinear Plasma Theory} (Academic Press, New York, 1972).
\bibitem{eawkp31} P. Sprangle, E. Esarey, and A. Ting, Phys. Rev. A \textbf{41}, 4463 (1990).
\bibitem{eawkp32} C. S. Kueny and P. J. Morrison, Phys. Plasmas \textbf{2}, 1926 (1995).
\bibitem{eawkp33} D. A. Klimachkov, A.S. Petrosyan, J. Exp. Theor. Phys. \textbf{122}, 832 (2016).

\bibitem{eawkp031} Y. -C. Ma,  Stud. Appl. Math. \textbf{60}(1), 43 (1979).
\bibitem{eawkp032} N. N. Akhmediev, V. M. Eleonskii, N. Kulagin,  Theor. Math. Phys. \textbf{72}(2), 809 (1987).
\bibitem{eawkp033} S. Tao, AIMS Math. \textbf{7}, 15795 (2022).
\bibitem{eawkp034} S. Nasipuri, P. Chatterjee, U. N. Ghosh, Eur. Phys. J. D \textbf{78}(7), 1 (2024)
\bibitem{eawkp035} P. Chatterjee, L. Mandi, and J. Mondal, Contrib. Plasma Phys., e70067 (2025), https://doi.org/10.1002/ctpp.70067.
\bibitem{eawkp036} S. V. Manakov, Vl. E. Zakharov, L.A. Bordag, A. R. Its, V. B. Matveev, Phys. Lett. A  \textbf{63}, 205 (1977).

\bibitem{eawkp34} Y. Nakamura, H. Bailung, and K. E. Lonngren, Phys. Plasmas \textbf{6}, 3466 (1999).
\bibitem{eawkp35} S. C. Tsang, K S. Chiang, K. W. Chow, Opt. Commun. \textbf{229}, 431 (2004).
\bibitem{eawkp36} H. Demiray, Appl Math Lett. \textbf{18}, 941 (2005).
\bibitem{eawkp37} C. H. Su and R. M. Mirie, J. Fluid Mech. \textbf{98}, 509 (1980).
\bibitem{eawkp38} C. S. Gardner, J. M. Greener, M. D. Kruskal, and R. M. Miura, Phys. Rev. Lett. \textbf{19}, (1967) 1095.
\bibitem{eawkp39} G. X. Huang and M.G. Velarde, Phys. Rev. E \textbf{53}, 2988 (1996).
\bibitem{eawkp40} J. -K. Xue, Phys. Rev. E \textbf{69}, 016403 (2004).
\bibitem{eawkp41} J. -K. Xue, Phys. Lett. A \textbf{331}, 409 (2004).
\bibitem{eawkp42} T. Tsuboi, Phys. Rev. A \textbf{40}, 2753 (1989).
\bibitem{eawkp43} S. Li and W. Duan, Eur. Phys. J. B \textbf{62}, 485 (2008).
\bibitem{eawkp44} F. Verheest, M. A. Hellberg, and W. A. Hereman, Phys. Rev. E \textbf{86}, 036402 (2012).
\bibitem{eawkp45} U. N. Ghosh, P. Chatterjee, and R. Roychoudhury, Phys. Plasmas \textbf{19}, 012113 (2012).
\bibitem{eawkp46} N. Saini and K. Singh, Phys. Plasmas \textbf{23}, 103701 (2016).
\bibitem{eawkp47} U. Imon and M. S. Alam, Waves in Random and Complex Media, (2023), https://doi.org/10.1080/17455030.2023.2280910.
\bibitem{eawkp48} A. Abdikian, U. N. Ghosh, and M. Eghbali, Braz. J. Phys. \textbf{55}, 27 (2025).
\bibitem{eawkp49} P. Eslami, M. Mottaghizadeh, and H. R. Pakzad, Astrophys. Space Sci. \textbf{338}, 271 (2012).
\bibitem{eawkp50} R. Jahangir and W. Masood, Phys. Plasmas \textbf{27}, 042105 (2020).
\bibitem{eawkp51} S. Akter and M. G. Hafez, AIP Advances \textbf{13}, 015005 (2023).

\bibitem{kp3} N. Batool, W. Masood, M. A. Huwayz, A. H. Almuqrin, and S. A. El-Tantawy, Phys. Plasmas \textbf{32}, 042102 (2025).
\bibitem{kp4} M. K. Ghorui, U. K. Samanta, P. Chatterjee, Astrophys Space Sci.  \textbf{345}, 273 (2013).
\bibitem{kp5} L. Mandi1, J. Mondal, P. Chatterjee, and S. Raut, Eur. Phys. J. D \textbf{79}, 102 (2025).

\bibitem{kp6} S. Nasipuri, S. Chandra, U.N. Ghosh, C. Das, P. Chatterjee, Indian J. Phys. \textbf{99}, 4389 (2025).
\bibitem{kp7} S. Akther, M. G. Hafez, Alexandria Engineering Journal \textbf{122}, 520–532(2025).
\bibitem{kp8} P. Schippers, M. Blanc, N. Andre, I. Dandouras, G. R. Lewis, L. K. Gilbert, A. M. Persoon, N. Krupp, D. A. Gurnett, A. J. Coates, S. M. Krimigis, D. T. Young, and M. K. Dougherty, J. Geophys. Res. \textbf{113}, A07208 (2008).

\end{thebibliography}
\end{document}